\documentclass[12pt,a4paper,titlepage]{article}
\usepackage[utf8]{inputenc}
\usepackage[top=50pt,bottom=50pt,left=68pt,right=66pt]{geometry}
\usepackage{amsmath}
\usepackage{booktabs}
\usepackage{tikz}
\usetikzlibrary{arrows, calc, positioning}
\usetikzlibrary{arrows.meta}
\usepackage{amssymb}
\usepackage[title]{appendix}
\usepackage{mathrsfs}
\usepackage{float}
\usepackage{multirow}
\usepackage{graphicx,caption,subcaption}
\usepackage[space]{grffile}
\usepackage{cite}

\begin{document}
\renewcommand{\arraystretch}{1.3}

\makeatletter
\def\@hangfrom#1{\setbox\@tempboxa\hbox{{#1}}%
      \hangindent 0pt%\wd\@tempboxa
      \noindent\box\@tempboxa}
\makeatother

% Underline for text or math

\def\un#1{\relax\ifmmode\@@underline#1\else
        $\@@underline{\hbox{#1}}$\relax\fi}

% Accents and foreign (in text):

\let\under=\unt                 % bar-under (but see \un above)
\let\ced=\ce                    % cedilla
\let\du=\du                     % dot-under
\let\um=\Hu                     % Hungarian umlaut
\let\sll=\lp                    % slashed (suppressed) l (Polish)
\let\Sll=\Lp                    % " L
\let\slo=\os                    % slashed o (Scandinavian)
\let\Slo=\Os                    % " O
\let\tie=\ta                    % tie-after (semicircle connecting two letters)
\let\br=\ub                     % breve
                % Also: \`        grave
                %       \'        acute
                %       \v        hacek (check)
                %       \^        circumflex (hat)
                %       \~        tilde (squiggle)
                %       \=        macron (bar-over)
                %       \.        dot (over)
                %       \"        umlaut (dieresis)
                %       \aa \AA   A-with-circle (Scandinavian)
                %       \ae \AE   ligature (Latin & Scandinavian)
                %       \oe \OE   " (French)
                %       \ss       es-zet (German sharp s)
                %       \$  \#  \&  \%  \pounds  {\it\&}  \dots

% Abbreviations for Greek letters

\def\a{\alpha}
\def\b{\beta}
\def\c{\chi}
\def\d{\delta}
\def\e{\epsilon}
\def\f{\phi}
\def\g{\gamma}
\def\h{\eta}
\def\i{\iota}
\def\j{\psi}
\def\k{\kappa}
\def\l{\lambda}
\def\m{\mu}
\def\n{\nu}
\def\o{\omega}
\def\p{\pi}
\def\q{\theta}
\def\r{\rho}
\def\s{\sigma}
\def\t{\tau}
\def\u{\upsilon}
\def\x{\xi}
\def\z{\zeta}
\def\D{\Delta}
\def\F{\Phi}
\def\G{\Gamma}
\def\J{\Psi}
\def\L{\Lambda}
\def\O{\Omega}
\def\P{\Pi}
\def\Q{\Theta}
\def\S{\Sigma}
\def\U{\Upsilon}
\def\X{\Xi}

% Varletters

\def\ve{\varepsilon}
\def\vf{\varphi}
\def\vr{\varrho}
\def\vs{\varsigma}
\def\vq{\vartheta}

% Calligraphic letters

\def\ca{{\cal A}}
\def\cb{{\cal B}}
\def\cc{{\cal C}}
\def\cd{{\cal D}}
\def\ce{{\cal E}}
\def\cf{{\cal F}}
\def\cg{{\cal G}}
\def\ch{{\cal H}}
\def\ci{{\cal I}}
\def\cj{{\cal J}}
\def\ck{{\cal K}}
\def\cl{{\cal L}}
\def\cm{{\cal M}}
\def\cn{{\cal N}}
\def\co{{\cal O}}
\def\cp{{\cal P}}
\def\cq{{\cal Q}}
\def\car{{\cal R}}
\def\cs{{\cal S}}
\def\ct{{\cal T}}
\def\cu{{\cal U}}
\def\cv{{\cal V}}
\def\cw{{\cal W}}
\def\cx{{\cal X}}
\def\cy{{\cal Y}}
\def\cz{{\cal Z}}

% Fonts

\def\Sc#1{{\hbox{\sc #1}}}      % script for single characters in equations
\def\Sf#1{{\hbox{\sf #1}}}      % sans serif for single characters in equations

                        % Also:  \rm      Roman (default for text)
                        %        \bf      boldface
                        %        \it      italic
                        %        \mit     math italic (default for equations)
                        %        \sl      slanted
                        %        \em      emphatic
                        %        \tt      typewriter
                        % and sizes:    \tiny
                        %               \scriptsize
                        %               \footnotesize
                        %               \small
                        %               \normalsize
                        %               \large
                        %               \Large
                        %               \LARGE
                        %               \huge
                        %               \Huge

% Math symbols

\def\slpa{\slash{\pa}}                            % slashed partial derivative
\def\slin{\SLLash{\in}}                                   % slashed in-sign
\def\bo{{\raise-.3ex\hbox{\large$\Box$}}}               % D'Alembertian
\def\cbo{\Sc [}                                         % curly "
\def\pa{\partial}                                       % curly d
\def\de{\nabla}                                         % del
\def\dell{\bigtriangledown}                             % hi ho the dairy-o
\def\su{\sum}                                           % summation
\def\pr{\prod}                                          % product
\def\iff{\leftrightarrow}                               % <-->
\def\conj{{\hbox{\large *}}}                            % complex conjugate
\def\ltap{\raisebox{-.4ex}{\rlap{$\sim$}} \raisebox{.4ex}{$<$}}   % < or ~
\def\gtap{\raisebox{-.4ex}{\rlap{$\sim$}} \raisebox{.4ex}{$>$}}   % > or ~
\def\TH{{\raise.2ex\hbox{$\displaystyle \bigodot$}\mskip-4.7mu \llap H \;}}
\def\face{{\raise.2ex\hbox{$\displaystyle \bigodot$}\mskip-2.2mu \llap {$\ddot
        \smile$}}}                                      % happy face
\def\dg{\sp\dagger}                                     % hermitian conjugate
\def\ddg{\sp\ddagger}                                   % double dagger
                        % Also:  \int  \oint              integral, contour
                        %        \hbar                    h bar
                        %        \infty                   infinity
                        %        \sqrt                    square root
                        %        \pm  \mp                 plus or minus
                        %        \cdot  \cdots            centered dot(s)
                        %        \oplus  \otimes          group theory
                        %        \equiv                   equivalence
                        %        \sim                     ~
                        %        \approx                  approximately =
                        %        \propto                  funny alpha
                        %        \ne                      not =
                        %        \le \ge                  < or = , > or =
                        %        \{  \}                   braces
                        %        \to  \gets               -> , <-
                        % and spaces:  \,  \:  \;  \quad  \qquad
                        %              \!                 (negative)

\font\tenex=cmex10 scaled 1200

% Math stuff with one argument

\def\sp#1{{}^{#1}}                              % superscript (unaligned)
\def\sb#1{{}_{#1}}                              % sub"
\def\oldsl#1{\rlap/#1}                          % poor slash
\def\slash#1{\rlap{\hbox{$\mskip 1 mu /$}}#1}      % good slash for lower case
\def\Slash#1{\rlap{\hbox{$\mskip 3 mu /$}}#1}      % " upper
\def\SLash#1{\rlap{\hbox{$\mskip 4.5 mu /$}}#1}    % " fat stuff (e.g., M)
\def\SLLash#1{\rlap{\hbox{$\mskip 6 mu /$}}#1}      % slash for no-in sign
\def\PMMM#1{\rlap{\hbox{$\mskip 2 mu | $}}#1}   %
\def\PMM#1{\rlap{\hbox{$\mskip 4 mu ~ \mid $}}#1}       %
\def\Tilde#1{\widetilde{#1}}                    % big tilde
\def\Hat#1{\widehat{#1}}                        % big hat
\def\Bar#1{\overline{#1}}                       % big bar
\def\sbar#1{\stackrel{*}{\Bar{#1}}}             % big bar with star
\def\bra#1{\left\langle #1\right|}              % < |
\def\ket#1{\left| #1\right\rangle}              % | >
\def\VEV#1{\left\langle #1\right\rangle}        % < >
\def\abs#1{\left| #1\right|}                    % | |
\def\leftrightarrowfill{$\mathsurround=0pt \mathord\leftarrow \mkern-6mu
        \cleaders\hbox{$\mkern-2mu \mathord- \mkern-2mu$}\hfill
        \mkern-6mu \mathord\rightarrow$}
\def\dvec#1{\vbox{\ialign{##\crcr
        \leftrightarrowfill\crcr\noalign{\kern-1pt\nointerlineskip}
        $\hfil\displaystyle{#1}\hfil$\crcr}}}           % <--> accent
\def\dt#1{{\buildrel {\hbox{\LARGE .}} \over {#1}}}     % dot-over for sp/sb
\def\dtt#1{{\buildrel \bullet \over {#1}}}              % alternate "
\def\der#1{{\pa \over \pa {#1}}}                % partial derivative
\def\fder#1{{\d \over \d {#1}}}                 % functional derivative
                % Also math accents:    \bar
                %                       \check
                %                       \hat
                %                       \tilde
                %                       \acute
                %                       \grave
                %                       \breve
                %                       \dot    (over)
                %                       \ddot   (umlaut)
                %                       \vec    (vector)

% Math stuff with more than one argument

\def\fracmm#1#2{{\textstyle{#1\over\vphantom2\smash{\raise.20ex
        \hbox{$\scriptstyle{#2}$}}}}}                   % fraction
\def\half{\frac12}                                        % 1/2
\def\sfrac#1#2{{\vphantom1\smash{\lower.5ex\hbox{\small$#1$}}\over
        \vphantom1\smash{\raise.4ex\hbox{\small$#2$}}}} % alternate fraction
\def\bfrac#1#2{{\vphantom1\smash{\lower.5ex\hbox{$#1$}}\over
        \vphantom1\smash{\raise.3ex\hbox{$#2$}}}}       % "
\def\afrac#1#2{{\vphantom1\smash{\lower.5ex\hbox{$#1$}}\over#2}}    % "
\def\partder#1#2{{\partial #1\over\partial #2}}   % partial derivative of
\def\parvar#1#2{{\d #1\over \d #2}}               % variation of
\def\secder#1#2#3{{\partial^2 #1\over\partial #2 \partial #3}}  % second "
\def\on#1#2{\mathop{\null#2}\limits^{#1}}               % arbitrary accent
\def\bvec#1{\on\leftarrow{#1}}                  % backward vector accent
\def\oover#1{\on\circ{#1}}                              % circle accent

\def\[{\lfloor{\hskip 0.35pt}\!\!\!\lceil}
\def\]{\rfloor{\hskip 0.35pt}\!\!\!\rceil}
\def\Lag{{\cal L}}
\def\du#1#2{_{#1}{}^{#2}}
\def\ud#1#2{^{#1}{}_{#2}}
\def\dud#1#2#3{_{#1}{}^{#2}{}_{#3}}
\def\udu#1#2#3{^{#1}{}_{#2}{}^{#3}}
\def\calD{{\cal D}}
\def\calM{{\cal M}}

\def\szet{{${\scriptstyle \b}$}}
\def\ulA{{\un A}}
\def\ulM{{\underline M}}
\def\cdm{{\Sc D}_{--}}
\def\cdp{{\Sc D}_{++}}
\def\vTheta{\check\Theta}
\def\fracm#1#2{\hbox{\large{${\frac{{#1}}{{#2}}}$}}}
\def\ha{{\fracmm12}}
\def\tr{{\rm tr}}
\def\Tr{{\rm Tr}}
\def\itrema{$\ddot{\scriptstyle 1}$}
\def\ula{{\underline a}} \def\ulb{{\underline b}} \def\ulc{{\underline c}}
\def\uld{{\underline d}} \def\ule{{\underline e}} \def\ulf{{\underline f}}
\def\ulg{{\underline g}}
\def\items#1{\\ \item{[#1]}}
\def\ul{\underline}
\def\un{\underline}
\def\fracmm#1#2{{{#1}\over{#2}}}
\def\footnotew#1{\footnote{\hsize=6.5in {#1}}}
\def\low#1{{\raise -3pt\hbox{${\hskip 0.75pt}\!_{#1}$}}}

\def\Dot#1{\buildrel{_{_{\hskip 0.01in}\bullet}}\over{#1}}
\def\dt#1{\Dot{#1}}

\def\DDot#1{\buildrel{_{_{\hskip 0.01in}\bullet\bullet}}\over{#1}}
\def\ddt#1{\DDot{#1}}

\def\DDDot#1{\buildrel{_{_{\hskip 0.01in}\bullet\bullet\bullet}}\over{#1}}
\def\dddt#1{\DDDot{#1}}

\def\DDDDot#1{\buildrel{_{_{\hskip 
0.01in}\bullet\bullet\bullet\bullet}}\over{#1}}
\def\ddddt#1{\DDDDot{#1}}

\def\Tilde#1{{\widetilde{#1}}\hskip 0.015in}
\def\Hat#1{\widehat{#1}}

% Aligned equations

\newskip\humongous \humongous=0pt plus 1000pt minus 1000pt
\def\caja{\mathsurround=0pt}
\def\eqalign#1{\,\vcenter{\openup2\jot \caja
        \ialign{\strut \hfil$\displaystyle{##}$&$
        \displaystyle{{}##}$\hfil\crcr#1\crcr}}\,}
\newif\ifdtup
\def\panorama{\global\dtuptrue \openup2\jot \caja
        \everycr{\noalign{\ifdtup \global\dtupfalse
        \vskip-\lineskiplimit \vskip\normallineskiplimit
        \else \penalty\interdisplaylinepenalty \fi}}}
\def\li#1{\panorama \tabskip=\humongous                         % eqalignno
        \halign to\displaywidth{\hfil$\displaystyle{##}$
        \tabskip=0pt&$\displaystyle{{}##}$\hfil
        \tabskip=\humongous&\llap{$##$}\tabskip=0pt
        \crcr#1\crcr}}
\def\eqalignnotwo#1{\panorama \tabskip=\humongous
        \halign to\displaywidth{\hfil$\displaystyle{##}$
        \tabskip=0pt&$\displaystyle{{}##}$
        \tabskip=0pt&$\displaystyle{{}##}$\hfil
        \tabskip=\humongous&\llap{$##$}\tabskip=0pt
        \crcr#1\crcr}}

% Some defs

\def\eV{\,{\rm eV}}
\def\keV{\,{\rm keV}}
\def\MeV{\,{\rm MeV}}
\def\GeV{\,{\rm GeV}}
\def\TeV{\,{\rm TeV}}
\def\sv{\left<\sigma v\right>}
\def\({\left(}
\def\){\right)}
\def\cm{{\,\rm cm}}
\def\K{{\,\rm K}}
\def\kpc{{\,\rm kpc}}
\def\beq{\begin{equation}}
\def\eeq{\end{equation}}
\def\bea{\begin{eqnarray}}
\def\eea{\end{eqnarray}}

% New commands

\newcommand{\be}{\begin{equation}}
\newcommand{\ee}{\end{equation}}
\newcommand{\nbe}{\begin{equation*}}
\newcommand{\nee}{\end{equation*}}

\newcommand{\fr}{\frac}
\newcommand{\lb}{\label}

\thispagestyle{empty}

{\hbox to\hsize{
\vbox{\noindent August 2025 \hfill IPMU25-0026 \\
\noindent  \hfill }}

\noindent
\vskip2.0cm
\begin{center}

{\large\bf Are single-field models of inflation and PBHs production ruled out by ACT observations?}

\vglue.3in

Daniel Frolovsky~${}^{a,*}$ and Sergei V. Ketov~${}^{a,b,c,\#}$ 
\vglue.3in

${}^a$~Department of Physics and Interdisciplinary Research Laboratory, \\
Tomsk State University, Tomsk 634050, Russia\\
${}^b$~Department of Physics, Tokyo Metropolitan University, Tokyo 192-0397, Japan \\
${}^c$~Kavli Institute for the Physics and Mathematics of the Universe (WPI),
\\The University of Tokyo Institutes for Advanced Study,  Chiba 277-8583, Japan\\
\vglue.2in

${}^{*}$~frolovsky@mail.tsu.ru, ${}^{\#}$~ketov@tmu.ac.jp
\end{center}

\vglue.3in

\begin{center}
{\Large\bf Abstract}

\end{center}

The data release from the Atacama Cosmology Telescope (ACT) imposes stronger constraints on primordial black holes (PBHs) formation in single-field inflation models  versus the Planck data. In particular, the updated Cosmic Microwave Background (CMB) radiation measurements favour a {\it higher} scalar spectral index $n_s$ and its {\it positive} running $\alpha_s$, which put the single-field models under scrutiny. Even in the absence of PBHs production, the new data constrain many single-field models of inflation. To explore this tension, we study PBHs formation in a concrete viable $\alpha$-attractor E-model. We investigate an impact of bending of the inflaton potential plateau toward reconciling the model with the ACT bounds on the CMB observables. We find that attempts to increase $n_s$ through bending lead to negative values of $\alpha_s$. Those values are disfavored by the ACT bounds just above $2\sigma$ even for PBHs in the asteroid-mass range, while the tension becomes stronger for heavier PBHs.

\vglue.2in
\noindent

\section{Introduction}

The paradigm of cosmic inflation in the early Universe provides an explanation of the observed properties of the Cosmic Microwave Background (CMB) radiation \cite{Planck:2018jri}. The increasing precision of CMB observations continues to drive the development of inflation models \cite{Dioguardi:2025vci, Kallosh:2025rni, Gialamas:2025ofz, Aoki:2025wld, Berera:2025vsu,Brahma:2025dio, Gialamas:2025kef, Salvio:2025izr, Antoniadis:2025pfa, Dioguardi:2025mpp, Kuralkar:2025zxr,Gao:2025onc,He:2025bli,Drees:2025ngb,Haque:2025uri,Liu:2025qca,Addazi:2025qra, Byrnes:2025kit, Yogesh:2025wak, Kallosh:2025ijd, Kohri:2025lau, Ahmad:2025mul, Okada:2025lpl, Choudhury:2025vso, Han:2025cwk, Leontaris:2025hly, McDonald:2025tfp, Gao:2025viy, Wolf:2025ecy }. Yet even in the simple case of single-field slow-roll inflation, CMB observations alone cannot uniquely determine the underlying model \cite{Karam:2022nym}. To solve the horizon and flatness problems, the inflaton potential must exhibit a plateau that extends for about $50-60$ e-folds. Such a plateau gives rise to an almost scale-invariant spectrum of scalar perturbations, whose tilts and amplitude must be consistent with observational constraints. 

The formation of PBHs from collapse of large scalar perturbations generated during inflation may offer additional insights into the underlying theory of inflation \cite{Novikov:1967tw, Hawking:1971ei, Carr:2016drx, Carr:2020gox}. Large perturbations may be generated via inflationary dynamics driven by localized features in the inflaton potential, such as a nearly-inflection point \cite{Ivanov:1994pa, Garcia-Bellido:1996mdl, Garcia-Bellido:2017mdw, Germani:2017bcs}. Then the power spectrum of scalar perturbations has a peak, whose position corresponds to the PBHs mass scale. A gravitational collapse of large scalar perturbations induces gravitational waves (GWs) that can be detected by current and future experiments, see Ref.~\cite{Domenech:2021ztg} for a review. It is possible to perform the reconstruction chain from an GW signal to the  scalar power spectrum and then to the  inflaton potential \cite{LISACosmologyWorkingGroup:2024hsc, Frolovsky:2024zds, LISACosmologyWorkingGroup:2025vdz, Frolovsky:2025qre}. 

However, in single-field inflation models, the features leading to PBHs generically lead to a lower value of the CMB spectral index $n_s$. Furthermore, the heavier those PBHs are, the lower the value of $n_s$ is, because a peak and a dip in the power spectrum are shifted toward the CMB scales and distort the plateau \cite{Frolovsky:2023xid}. For instance, the $\alpha$-attractor T-model modified to generate PBHs in the asteroid-mass range \cite{Dalianis:2018frf} becomes incompatible with the Planck 2018 constraints  \cite{Planck:2018jri}:
\begin{equation}
n_s = 0.9649 \pm 0.0042\quad (68\%\, \text{CL}),\quad  \alpha_s \equiv \mathrm{d}\,n_s/\mathrm{d} \ln k = -0.0045 \pm 0.0067 \quad  (68\%\, \text{CL}).
\end{equation}

The more recent ACT data release in combination with DESI and Planck data imposes the tighter constraints \cite{ACT:2025fju,ACT:2025tim,DESI:2024mwx}:
\begin{equation}\label{act}
n_s = 0.9743 \pm 0.0034\quad (68\%\, \text{CL}),\quad \alpha_s \equiv \mathrm{d}\,n_s/\mathrm{d} \ln k = 0.0062 \pm 0.0052 \quad (68\%\, \text{CL}).
\end{equation}
These  bounds present new challenges for single-field inflation models, especially  for those involving PBHs production. In single-field models, $n_s$ and $\alpha_s$ are often given in terms  e-folds $N_e$ as
\begin{equation}\label{nsalp}
	n_s = 1-\frac{q}{N_e}+{\cal O}(N^{-2}_e)~, \qquad \alpha_s\equiv -\,\mathrm{d}\,n_s / \mathrm{d}\, N_e = -  \,\frac{q}{N_e^2}+{\cal O}(N^{-3}_e)~,\end{equation}
where $q$ is a positive model-dependent constant. For instance, the standard Starobinsky inflation model \cite{Starobinsky:1980te, Ketov:2025nkr} has $q = 2$ leading to $n_s \approx 0.966$ and $\alpha_s \approx -0.0005$ for  $N_e=60$ that differs from the ACT+DESI+Planck best-fit.

Another example is a generalization of chaotic inflation with a non-minimal coupling to gravity, which was proposed in light of ACT data in Ref.~\cite{Kallosh:2025rni}. This model corresponds to $q = 3/2$ and predicts the value of $n_s$ consistent with current observational constraints for $N_e = 60$, while also yielding a negative running $\alpha_s \approx -0.0004$. 

Martin, Ringeval and Vennin in Ref.~\cite{Martin:2024nlo} analysed about 300 single-field inflation models and found that negative running of the scalar spectral index $\alpha_s$  is a common feature, with $\alpha_s = -6.3 \times 10^{-4}$ being the
most likely value. In the landscape of those models, positive values of $\alpha_s$ are excluded by $3\sigma$.

In this Letter, we explore theoretically motivated modifications of single-field inflation models to address the tension outlined above, and analyse their impact on the predicted CMB observables and the global structure of the inflaton potential by using the E-model of $\alpha$-attractors as an example.

\section{Bending Inflaton Potential}

A connection between the shape of an inflaton potential and its predictions to CMB observables becomes transparent in the slow-roll approximation with
\begin{equation}
	n_s=1+2\eta(\phi_*)-6\epsilon(\phi_*)\,, \qquad  r=16\epsilon(\phi_*)\,,
\end{equation}
where
\begin{equation}
	\epsilon(\phi)=\frac{M_{\mathrm{Pl}}^2}{2}\left(\frac{V^{\prime}(\phi)}{V(\phi)}\right)^2, \qquad \eta(\phi)=M_{\mathrm{Pl}}^2 \frac{V^{\prime \prime}(\phi)}{V(\phi)}~,
\end{equation}
are the standard slow-roll parameters, and $\phi_*$ is the value of the inflaton field at the horizon crossing on the standard pivot scale, $k=0.05\,{\rm Mpc}^{-1}$. As is clear from these equations, altering the spectral tilt $n_s$ and its running $\alpha_s$ requires modifying (bending) a plateau of the inflaton potential. This is more than just an ad hoc solution to the CMB-tension, while there are several theoretical reasons suggesting that the plateau in the inflaton potential cannot be arbitrarily long 
\cite{Ketov:2025nkr}.

As is well known, an inflation model based on modified $F(R)$-gravity can be transformed to the  standard (Einstein) gravity minimally coupled to a scalar field with the potential $V(\phi)$, whose shape is determined by the function $F(R)$. The higher-order curvature corrections beyond the $R^2$ term in $F(R)$ can easily spoil the flatness of the potential and bend the plateau, see e.g., \cite{Addazi:2025qra}.  Also, in string inflation, various perturbative corrections can alter the potential in a similar way \cite{Cicoli:2020bao, Leontaris:2025hly}. From a different perspective, if a de-Sitter spacetime is treated as a coherent quantum state of microscopic constituents, this state is subject to quantum depletion that imposes an upper limit on the possible total number of e-folds \cite{Dvali:2013eja,Dvali:2017eba}. In supergravity embeddings of inflation models  and their string theory realisations, inflaton $\phi$ can be interpreted as the dilaton field whose value is related to the volume of extra dimensions. Then the inflaton potential has the runaway behavior and asymptotically vanishes at large field values signalling a decompactification of the extra dimensions \cite{Dine:1985he,Alexandrov:2016plh,Frolovsky:2024xet}. There are other constraints on the length of inflation plateau, which follow from the Swampland Distance (SDC) and Trans-Planckian Censorship conjectures (TPC), which set an upper bound on the allowed inflaton field excursions $\D\phi$ in the effective field theory. The upper bound can be given in terms of the tensor-to-scalar ratio $r$ and the amplitude of scalar perturbations $A_s$ as \cite{Bedroya:2019tba, Scalisi:2018eaz}
\begin{equation} \lb{swamp}
 \abs{\D \phi} \leq M_{\rm Pl}\ln \left( \fracmm{M_{\rm Pl}}{H_{\rm inf}}\right)  \approx \fracmm{M_{\rm Pl}}{2}\ln \left(\fracmm{2}{\p^2A_sr}\right),
 \end{equation}
 where $H_{\rm inf}$ is the Hubble value during inflation. According to CMB measuremetns, $A_s \approx 2.1 \cdot 10^{-9}$ and $r \leq 0.032$ \cite{BICEP:2021xfz}, so that $\abs{\Delta\phi} \lesssim 10\, M_{\mathrm{Pl}}$, i.e. the duration of inflation and the length of the slow-roll plateau cannot exceed 100 e-folds.  

Demanding efficient (i.e. relevant to the current dark matter) PBHs production after inflation leads to further constraints. Let us consider the $\alpha$-attractor E-model of inflation with PBHs production at smaller scales,  which has the potential \cite{Frolovsky:2022qpg, Frolovsky:2023hqd}
\begin{equation} \label{epott1}
V(\phi) = \frac{3}{4}M^2M^2_{\rm Pl}\left[ 1 - y + \theta y^{-2} + y^2(\beta - \gamma y)\right]^2,  \qquad y = \exp \left( -\sqrt{\frac{2}{3\alpha}} \f/M_{\rm Pl}\right),
\end{equation}
where $M$ is the inflaton (Starobinsky) mass of the order $10^{13}$ GeV, and $\alpha, \beta, \gamma, \theta$ are the dimensionless parameters.  

\begin{figure*}[h]
\centering
  \includegraphics[width=0.49\textwidth]{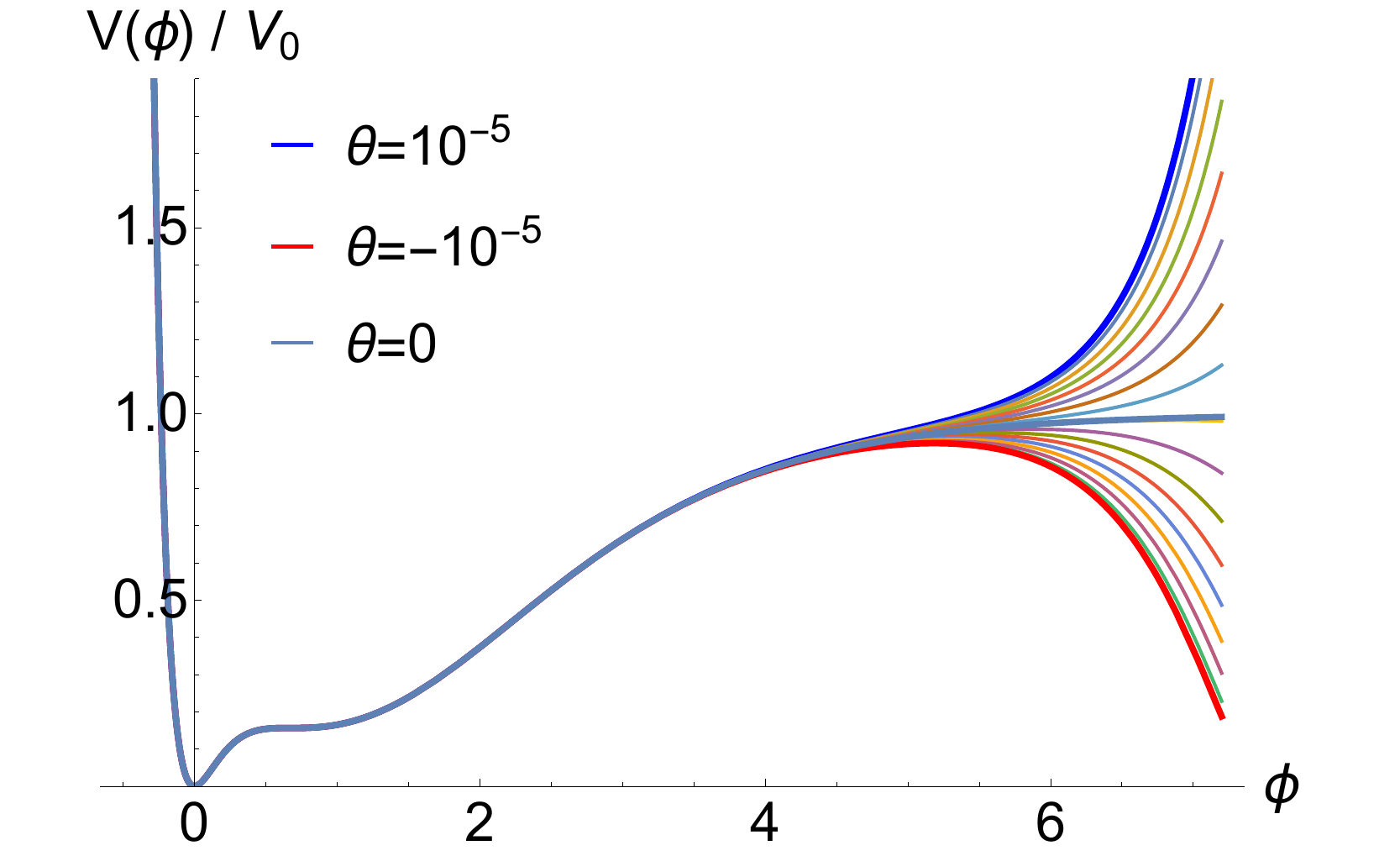}
  \caption{The potential in the E-model for various values of $\theta$ of the order  $10^{-5}$. The other parameters are tuned to generate PBHs with masses of the order $10^{19}\,\mathrm{g}$.}
  \label{vns}
\end{figure*}

To study an impact of bending the potential via the key parameter $\theta$ on the model predictions to CMB, we fix the other parameters to ensure PBHs production in the asteroid-mass  range. For details about the parameter selection and their interpretation, see Refs.~\cite{Frolovsky:2025qre, Frolovsky:2023hqd}.

\begin{figure*}[h]
  \includegraphics[width=0.49\textwidth]{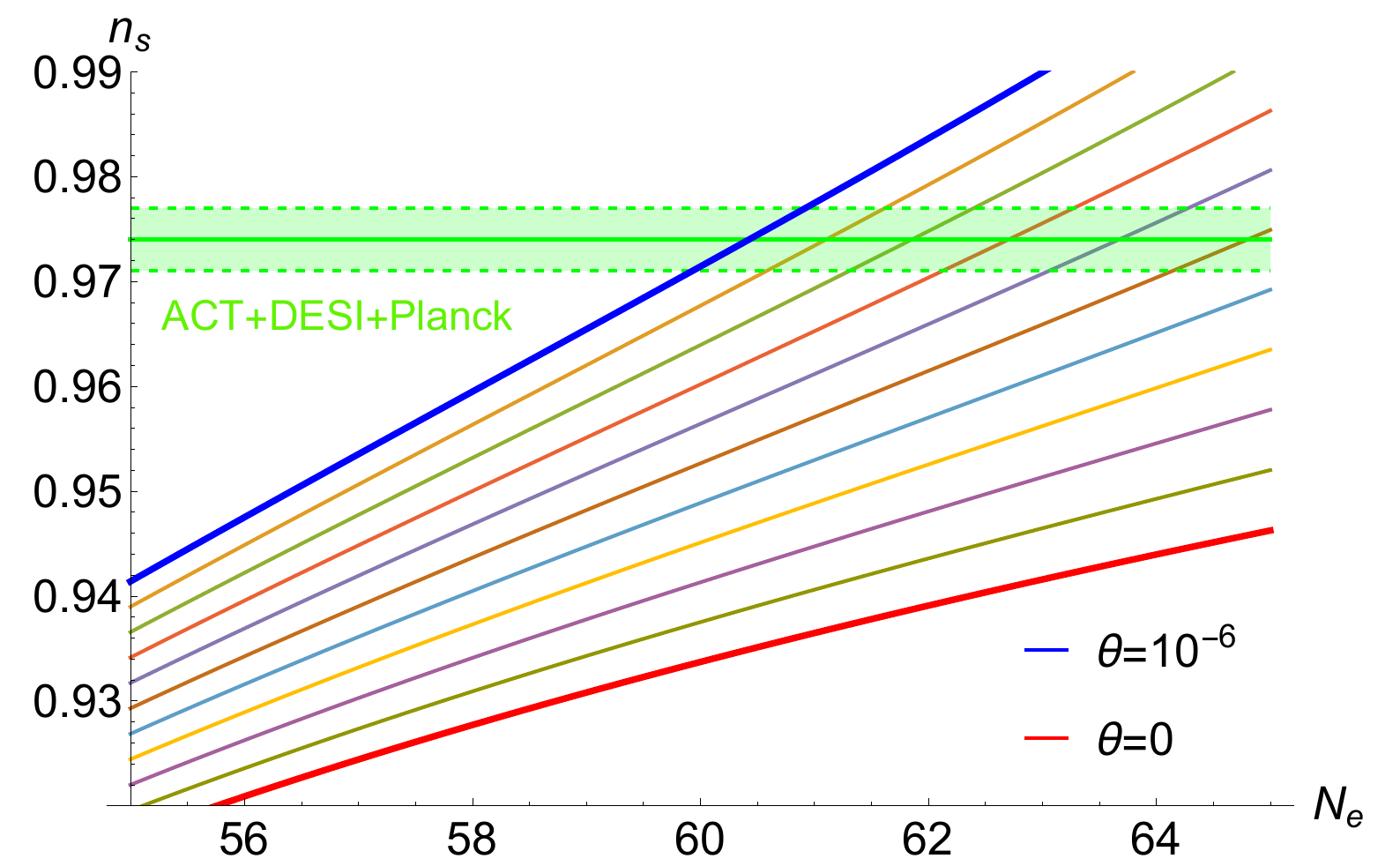}
  \includegraphics[width=0.49\textwidth]{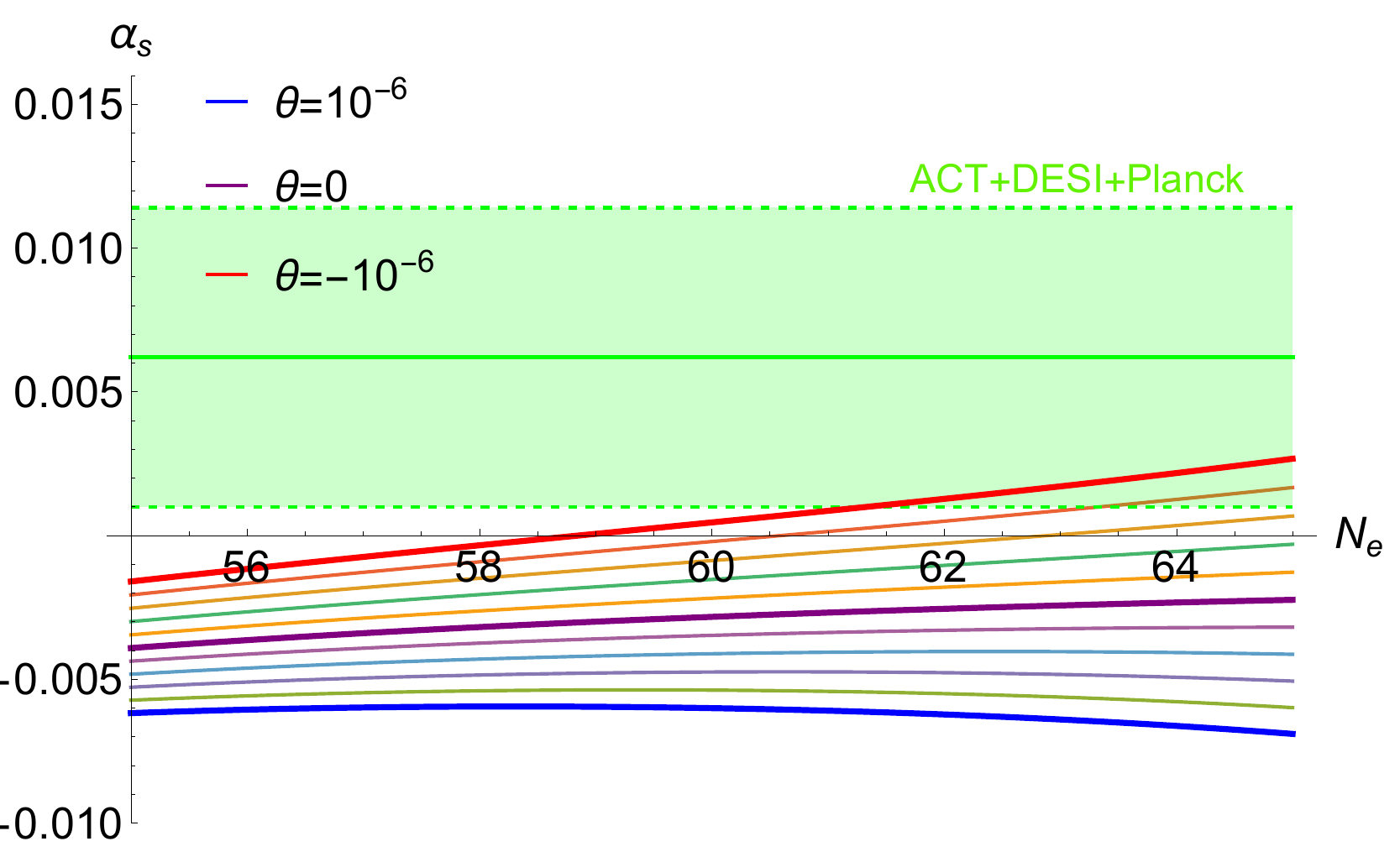}
  \caption{The dependence of $n_s$ upon $\theta$ in $[0, \ldots, 10^{-6}]$ (left), and $\alpha_s$ upon $\theta$ in $[-10^{-6}, \ldots, 10^{-6}]$ and e-folds $N_e$ (right). The other parameters are tuned to generate PBHs with masses of the order $10^{19}\,\mathrm{g}$. }
  \label{vns}
\end{figure*}

Figures 1 and 2 demonstrate that the $y^{-2}$ term in the potential \eqref{epott1} must have a {\it positive } coefficient in order to bend the plateau upward and thereby increase $n_s$. However, such upward bending also leads to a  {\it negative } value of $\alpha_s$, which is disfavoured by the latest ACT results. Moreover, the heavier the PBHs are, the larger a positive $\theta$ should be in order to match the observed $n_s$, and, hence, the stronger the tension against $\alpha_s$ is. 

A way to resolve this tension is to begin with a model that has a low positive $q \lesssim 1$ in Eq.~\eqref{nsalp} with the corresponding $n_s$ above the observational bound, and then bend the plateau downward to obtain a positive $\alpha_s$ and bring $n_s$ back to the allowed range.~\footnote{See Ref.~\cite{Leontaris:2025hly} for the example of successful bending in a single-field inflation model without PBHs production, which satisfies the ACT constraints  by using the potential of Eq.~(3.32) with five parameters.}  Otherwise, addressing this tension  requires more fine-tuning and imposes additional constraints on the initial conditions for the inflaton field.

\section{Discussion}

The ACT results combined with DESI and Planck data impose tighter constraints on the production of PBHs in single-field inflation models. Reconciling those models with the higher values of the spectral index $n_s$ requires upward bending of the inflaton potential plateau. However, obtaining a positive value of the running $\alpha_s$ simultaneously demands downward bending. Resolving this tension may require additional parameters and more fine-tuning. This sharpens the issue of  compatibility of PBHs formation from single-field inflation against the standard cosmology \cite{Allegrini:2024ooy}. On the other hand, stronger observational constraints make such models more predictive and more testable in the near future.

In this paper, we demonstrated that the $\alpha$-attractor E-model of inflation with PBHs production in the predicted asteroid-mass range is in tension with the ACT observations \cite{ACT:2025fju,ACT:2025tim,DESI:2024mwx}. The model predicts the negative value of $\alpha_s \approx -0.006$ with the PBHs masses being $\sim 10^{19},\mathrm{g}$, which is consistent with the central value from the Planck data  but is disfavored by ACT near the boundary  of the $2\sigma$ confidence region. The tension becomes stronger for heavier PBHs. We expect similar tension in other single-field models of inflation with PBHs production, see e.g.,  Ref.\cite{Allegrini:2024ooy}.  

Multi-field models can cure the tension but have less predictive power. In multi-field models, the peak in the scalar power spectrum typically has the log-normal shape \cite{Pi:2020otn}, unlike the broken power-law shape that is characteristic for single-field models. Many multi-field models do not have a dip before the peak, so they also avoid decreasing of $n_s$ that usually happens in single-field models \cite{Frolovsky:2023xid}. For instance, as was shown in Ref.~\cite{Kohri:2025lau}, introducing an additional scalar field can resolve the $(n_s,\alpha_s)$ tension, with the PBHs masses being  in the asteroid-mass range too.  

A quantification of the tension depends upon which combination of the available data is taken. There is also a 2$\sigma$ tension between BAO and CMB data within the standard cosmological model, as noted in Ref.~\cite{Ferreira:2025lrd}. Because of that, the constraints on inflationary models with PBHs production may be relaxed or tightened, depending on a chosen data set.

Future measurements of the tensor-to-scalar ratio by experiments such as LiteBIRD \cite{LiteBIRD:2022cnt}, Simons Observatory \cite{SimonsObservatory:2025wwn}, as well as the upcoming space-based gravitational wave interferometers LISA \cite{LISA:2017pwj}, TAIJI \cite{Gong:2014mca}, TianQin \cite{TianQin:2015yph} and DECIGO \cite{Kudoh:2005as} will provide complementary tests of the inflation scenarios involving PBHs formation. In the event of a gravitational wave detection caused by PBHs production, it may be possible to reconstruct the scalar power spectrum responsible for the signal. The reconstructed spectrum is supposed to match CMB observations on large scales, which is non-trivial in simple single-field models. This may provide the framework to test compatibility of the reconstructed inflaton potential with the underlying fundamental physics via supergravity, swampland conjectures, string theory and other quantum gravity considerations.

\section*{Acknowledgements}

The authors thank Michele Cicoli, Kazunori Kohri and William Wolf for correspondence.

DF and SVK were partially supported by Tomsk State University under the development program Priority-2030. DF was supported by the Foundation for Advancement of Theoretical Physics and Mathematics "BASIS".  SVK was also supported by Tokyo Metropolitan University and the World Premier International Research Center Initiative, MEXT, Japan.

\bibliographystyle{utphys}
\bibliography{Bibliography.bib}

\end{document}